\documentstyle[aps,preprint]{revtex}
\title{A New Perturbation Theory of\\ Finite-Size Effects near Critical Point}
\author{C.B. Yang$^{1,2}$ and X. Cai$^1$}
\address{
$^1$CCAST(World Laboratory), P.O. Box 8730, Beijing 100080, China}
\address{
$^2$Institute of Particle Physics, Hua-Zhong Normal University,
Wuhan 430070, China\footnote{Mailing address}}
\date{}
\draft
\begin{document}
\maketitle

\begin{abstract}
A new perturbation theory is proposed for studying finite-size effects near
critical point of the $\phi^4$ model with a one-component order parameter.
The new approach is based on the techniques of generating functional and
functional derivative with respect to external source field
and can be used for temperatures both above and below the critical
point of the bulk system. It is shown that this approach is much simpler
comparing with available perturbation
theories. Particularly, this new method avoids renormalization in calculating
many physical quantities such as correlation functions etc..

\end{abstract}
\vskip 1cm

Finite-size effects near critical points have been remained over the past two
decades to be an important topic of the active research both theoretically
and experimentally [1]. When one or more dimensions of a bulk
system is reduced to near or below a certain characteristic length scale,
the associated properties are modified reflecting the lower dimensionality.
It is believed that finite-size effects are precursors of the critical
behavior of the infinite system and can be exploited to extract the limiting
behavior. A central role plays the finite-size scaling behavior predicted by
both the phenomenological [2] and renormalization group [3] theories. Those
theories allowed a systematic discussion of the finite-size effects and,
consequently, form the cornerstone of our current understanding of the way in
which the singularities of an infinite system are modified by the finiteness
of the system in some or all of the dimensions. Of course, the exact form of
scaling functions can't be given in those scaling theories.

In 1985, Br\'{e}zin and Zinn-Justin (BZ) [4] and Rudnick, Guo and Jasnow
(RGJ) [5] developed two field-theoretical perturbation theories for the
calculation of the finite-size scaling functions within the $\phi^4$ model
which corresponds to the Ising model. Most applications of these theories to
three-dimensional systems have been restricted to region $T\geq T_C$ [6]
with a few calculations in region below $T_C$ [7]. In recent years the $\phi^4$
and the extended $\phi^6$ models have been used to investigate the multiplicity
fluctuations in the final states for first- and second-order phase transitions
of quark gluon plasma [8], under the approximation similar to the so-called
zero-mode approximation. As pointed out in the first paper in
Ref. [9], the theory of BZ is not applicable for $T<T_C$ and the results from
RGJ theory are not quantitatively reliable in the same temperature region
since the coefficients of the Gaussian terms in the integrals are negative
for those temperatures. Although the modified perturbation method in Ref. [9]
can be used for both $T>T_C$ and $T<T_C$, the calculation is lengthy and can
be done only to the first-order in practice. Since one does not know the exact
order of values of higher order terms, theoretical results have large
uncertainty. Furthermore, in the calculations of Ref. [4-7,9] there are
divergent terms which must be treated using renormalization technique which
not only is complicated but also increases theoretical uncertainty. Therefore,
a further study of the perturbation theory of finite-size effect is necessary.

It should be pointed out that all perturbation theories mentioned above are
based on the Fourier decomposition of the order parameter. This method is
natural because the decomposition enables one to transform the functional
integral into an infinite product of tractable normal integrals. Although
such decomposition has simple physical explanation which is very fruitful
for the understanding of properties of infinite systems and can deduce
reliable physical results, as in the case of usual field theories in particle
physics, it brings about a great deal of calculations for systems with
finite-size. This is not
surprising. As is well-known, quantities complicated in coordinate space may
have simple momentum spectra thus looks simple in momentum space, but those
obviously nonzero only in a finite range must have puzzling momentum spectra.
Therefore, for the study of properties of finite-size systems, calculations
in coordinate space might be simpler and more effective. The point here
is that one must calculate the complicated functional integral which is very
difficult to be evaluated directly.

In this Letter, we employ the technique of generating functional with the
help of external source field to develop a perturbation theory for the
study of the finite-size effects of systems near their critical points.
The external source field enables one
to turn the functional integral into much easier functional derivatives,
thus facilitates the calculations greatly. As will be shown, the perturbation
expansion can be carried out easily and one can get expressions for
physical quantities to second-order with little exertion.
More importantly, such technique isolates all the divergence in every order
of the expansion for the partition function into a common factor, showing the
common source of those divergences in the calculating of thermodynamical
quantities. Because of this factorization the renormalization can be done only
once for the perturbative calculations so that the expansion can be done to
high order needed with special difficulties. This advantage will make more
detailed investigation of finite-size effects possible.

In a $\phi^4$ model of phase transition with a one-component order parameter,
the partition function which is most important for the calculation of other
thermodynamical quantities can be expressed as a functional
integral of exponential of the Hamiltonian $H$ of the system
\begin{equation}
Z=\int {\cal D}\,\phi\,\exp(-H)=\int {\cal D}\,\phi\,\exp\left\{
-\int\,{\rm d}^d\,x\left[{\gamma\over 2}\phi^2+\frac{1}{2}(\nabla
\phi)^2+{u\over 4!}\phi^4\right]\right\}\ ,
\end{equation}

\noindent in which $\gamma=a(T-T_C)$, $a$ and $u$ are temperature dependent
positive constants, $T_C$ is the bulk critical temperature,
$\phi$ is the order parameter of the system, and the integration over $x$
is in the range of the finite volume.

Let's first consider the calculation of partition function in the region
$T>T_C$ i.e. $\gamma>0$. To calculate the functional integral in Eq. (1),
one can begin with considering a Gaussian functional integral with the
introduction of external source field $J$
\begin{eqnarray}
W[J]&=&\int{\cal D}\phi\exp\left\{-\int\,{\rm d}^d\,x\left[{\gamma\over 2}
\phi^2+\frac{1}{2}(\nabla\phi)^2-J\phi\right]\right\}
\nonumber\\
&=&\int{\cal D}\phi\exp\left\{-\int\,{\rm d}^d\,x\left[{\gamma\over 2}
\phi^2-\frac{1}{2}\phi\nabla^2\phi-J\phi\right]\right\}\ ,
\end{eqnarray}

\noindent where it is assumed that the $(d-1)$-dimensional surface integral
of $\phi\nabla\phi$ equals to zero which is satisfied for the usually assumed
periodic, anti-periodic, and Dirichlet boundary conditions. Following the
same standard procedures as in field theory for infinite system, one has
\begin{equation}
W[J]=\left[{\rm det}(\gamma-\nabla^2)\right]^{-1/2}\,\exp\left[
\frac{1}{2}\int{\rm d}^d\,x\,{\rm d}^d\,y\,J(x)\Delta(x,y)J(y)\right]\ ,
\end{equation}

\noindent where det represents the determinant of the operator
$\gamma-\nabla^2$ and the Green's function $\Delta(x,y)$ is the inverse of the
same operator, namely $\Delta(x, y)$ satisfies
\begin{equation}
(\gamma-\nabla^2)\Delta(x,y)=\delta(x-y)\ .
\end{equation}

\noindent  The boundary condition for $\Delta(x,y)$ can be chosen to be
Dirichlet type. This Green's function can be obtained analytically for some
systems with regular boundaries. For the case of the simplest
one-dimensional system within (0,L), one can show with a few algorithms
that $\Delta(x, y)$ is quite simple and can be written as
\begin{equation}
\Delta(x,y)=\cases{
{\rm sinh}\, \omega(L-y)\,{\rm sinh}\, \omega x/(\omega\,{\rm sinh}\,\omega L)
\ \ \ \ \ \ & \mbox{for $x<y$,}\cr
{\rm sinh}\, \omega(L-x)\,{\rm sinh}\, \omega y/ (\omega\,
{\rm sinh}\,\omega L)
\ \ \ \ \ \ & \mbox{for $x>y$\ ,}\cr  }
\end{equation}

\noindent in which $\omega=\sqrt{\gamma}$, ${\rm sinh}\,x\equiv
\frac{1}{2}(\exp(x)-\exp(-x))$ is the hyperbolic sine function.

Different from those for infinite systems, this Green's function for
finite-size system can't be written as a function of single variable $x-y$
even in the case with periodic boundary condition. The usually assumed
translational invariance is destroyed generally due to of the finite size
of the system and the system is invariant only under a subgroup of the
translational transformation under the assumption of periodic boundary
condition. It should also be pointed out that the last expression for the
Green's function can be continued to the temperature $T=T_C$, so that the
Green's function for $\gamma=0$ will be treated as the limiting of
$\gamma\to 0_+$ and will not be discussed in this Letter.

In terms of $W[J]$, the partition function $Z$ can be expressed as
\begin{eqnarray}
Z&=&\left.\exp\left(-\frac{u}{4!}\int{\rm d}^d\,x {\delta^4\over \delta
J^4(x)}\right)\,W[J]\, \right|_{J=0}\nonumber\\
&=&\left.\sum_{n=0}^{\infty}{(-u)^n\over n!(4!)^n}\left[\int{\rm
d}^d\,x{\delta^4\over\delta J^4(x)}\right]^n W[J]\, \right|_{J=0}
\end{eqnarray}

\noindent  Owing to the fact that it involves only functional derivatives,
the last expansion can be evaluated quite easily with the Wick's theorem.
In traditional perturbation theories, $Z$ can only be obtained up to
the first-order. In our new approach, the partition function can be calculated
to higher order and, for example, takes the form up to the second-order
approximation,
\begin{eqnarray}
Z&=&\left[{\rm det}(\gamma-\nabla^2)\right]^{-1/2}\,\left[1-\frac{u}{8}\int
{\rm d}^d\,x\,\Delta^2(x,x)+\frac{u^2}{8}\int {\rm d}^d\,x\,{\rm d}^d\,y
\right.\nonumber\\
&& \left.\left({\Delta^2(x,x) \Delta^2(y,y)\over 8}+\Delta(x,x)
\Delta^2(x,y)\Delta(y,y)+{\Delta^4(x,y)\over 3}\right)\right]\ .
\end{eqnarray}

\noindent Because no singularity is associated with the Green's function
$\Delta(x, y)$, a normal integral will give the result. In some cases for which
the Green's function has a simple form, above integral can be carried out
analytically. It should be noted that all terms in the perturbation expansion
are finite except a common factor $\left[{\rm det}(\gamma-\nabla^2)
\right]^{-1/2}$. This factor depends on the boundary condition and may
contribute to thermodynamical quantities such as entropy and heat capacity
etc.. The factor is generally divergent, therefore renormalization is needed
to obtain physically acceptable results. The renormalization for the factor
can be done with $\epsilon$-expansion and/or in fixed diemnsions, as shown
in Ref. [4-7,9]. So we do not discuss the renormalization of the factor
in this Letter. Since all the divergences have been isolated into the common
factor, the renormalization needs to be done only once, so that one can
expand the series to any order he needs with little effort. In fact, to every
next expansion order, the terms are accompanied with integration of two more
Green's functions over the volume of the system. As can be seen from an example
of the expression for the Green's function in one-dimensional case, The
integral of the two more Green's function is quite small, thus the first a
few terms in the expansion is enough in most cases, though higher order
calculations are not very difficult. More interestingly, the common divergent
factor is cancelled and needs not to be worried about when one is interested
in non-thermodynamical quantities such as correlation functions and moments
of the order parameter, leaving those physical quantities finite at any order
of the perturbation expansion. It is clearly seen from the symmetry property
of the Hamiltonian that correlation functions involving odd number points are
always zero so that only correlation functions involving even number points
are nonzero and finite. For illustration, the two-point correlation
function is calculated up to first-order approximation
\begin{equation}
\langle\phi(x)\phi(y)\rangle = \Delta(x,y)-\frac{u}{4}\int {\rm d}^d\,z
\Delta(x,z)\Delta(z,z)\Delta(z,y)\ .
\end{equation}

\noindent  One can see that no singularity exists in last expression and
its extension to higher orders is straightforward. The same is true for
other non-thermodynamical quantities such as moments of the order parameter
etc.. Due to the violation of translational invariance, the two-point
correlation function depends the coordinates $x,\ y$ of the two points
separately instead on their difference $x-y$. That is to say, $\langle
\phi(x)\phi(y)\rangle\neq\langle\phi(x-y)\phi(0)\rangle$, contrary to the
usual assumption of translational invariance, except in the very central
region of the system. This is different from traditional perturbation
theories for finite-size systems.

From above discussions one sees that the forms of the expansion are the same
for systems in one-, two- and three-dimensional spaces. To deal with spaces
with different dimensions, one needs only to use a new Green's function in
proper dimensional space and retains the form of the expansion. More
remarkably, if the shape of the boundary of the system is not a square nor
cubic, the Fourier decomposition can't be performed and the traditional
perturbation theories fail to work. This disadvantage limits the
application of traditional perturbation theories to realistic systems for
which the boundaries are usually not so regular. For these systems our new
approach can nevertheless be carried out because the needed Green's function
and the determinant of the operator $\gamma-\nabla^2$ can still be determined
numerically while the form of our perturbation expansion needs not to be
altered. This advantage will offer the possibility for one to study more
realistic models. It is certain that the comparison between results from the
new perturbation theory and real experiments can be done in the future.

When one intends to calculate quantities in the temperature region lower
than $T_C$, above procedures should be modified. In that temperature region
$\gamma$ is negative so that the functional integral in Eq. (2) is
ill-defined and can't be used in the calculations of the partition function
$Z$ and other physical quantities. In fact, this is the very origin of all
the difficulties in previous perturbation theories. The cause for the
difficulties can be understood physically. For $T>T_C$, the potential
$(\gamma/2)\phi^2+(u/4!)\phi^4$ takes its minimum at $\phi=0$
and fluctuations about the minimum should be small and can be taken into
account by perturbation theory with the help of $W[J]$ as shown above.
For $T<T_C$, however, the potential takes its extremum at $\phi=0$ and its
degenerate minima at $\phi=\pm\phi_0=\pm \sqrt{6a(T_C-T)/u}$\ . These
extrema can be easily understood by drawing the potential for this case.
That is to say, the most probable order parameter or ground state of the
system for temperature below $T_C$ is either $\phi_0$ or $-\phi_0$. The
ground states do not possess the symmetry $\phi\leftrightarrow-\phi$ of
the Lagrangian, or the symmetry is broken. This is known as the spontaneous
symmetry breaking which is very familiar in superconductors, crystal
lattices, and the buckling of a compressed needle. In such case,
$\langle \phi\rangle$ is nonzero but lies in the vicinity of either $\phi_0$
or $-\phi_0$. If the perturbation expansion is nevertheless be done around
$\phi=0$, it is bound to be very complicated and even to fail. Then we are
authorized to expand perturbation theory around one of the degenerate
classical minima of the potential. Keeping all above facts in mind, one can
shift the field to the vicinity of one of the minima through, for example,
\begin{equation}
\phi=\phi_0+\varphi
\end{equation}

\noindent and considers fluctuations of $\varphi$ around zero instead of those
of $\phi$. Then the Hamiltonian of the system takes the form
\begin{equation}
H=\int {\rm d}^d\,x\,\left[-{3\gamma^2\over 2u}+\frac{-\gamma}{2}\varphi^2
+\frac{1}{2}(\nabla\varphi)^2+\frac{u\phi_0}{6}\varphi^3+\frac{u}{4!}
\varphi^4\right] \ .
\end{equation}

\noindent The constant term in $H$ corresponds to the mean field result of
its free-energy. After the shift the coefficient $-\gamma/2$ of the Gaussian
terms in $H$ is positive. Now the partition function and other physical
quantities can be calculated in a similar way to that for the case $T>T_C$.
The new $\varphi^3$ term in $H$ shows the difference between the two phases
in the transition. This term has no contribution to the partition function
and two-point correlation function at first-order of the expansion, thus
those quantities depend on $|\gamma|$ at first-order approximation. Due to
the $\varphi^3$ term in $H$, the correlation functions among $2n+1$ points,
which are zero for high temperature phase, turn out to be nonzero up to the
first-order approximation.

Above discussions allow a systematic study of the critical properties of
finite-size systems with a one-component order parameter. Clearly, the
techniques can be generalized to the case with $n$-component order parameter,
which will be discussed elsewhere.

As a summary, a new perturbation theory of $\phi^4$ model with a
one-component order parameter is proposed to study finite-size effects of
system near critical point. The new method provides a simple way for the
calculation of the partition functions and correlation functions etc., and
can be generalized to more complicated and realistic models.

This work was supported in part by the NNSF, the Hubei SF and the SECF in
China.

\end{document}